**Title** *(160 characters)*

Protection of hamsters from mortality by reducing fecal moxifloxacin concentration with DAV131A in a model of moxifloxacin-induced *Clostridium difficile* colitis

# Authors

Charles Burdet[*], MD, MPH (1,2), Sakina Sayah-Jeanne[*], PhD (3), Thu Thuy Nguyen, PhD (1), Christine Miossec[§], PhD (3), Nathalie Saint-Lu, PhD (3), Mark Pulse, MS (4), William Weiss, MS (4), Antoine Andremont, MD, PhD (1,2), France Mentré[$], MD, PhD (1,2), Jean de Gunzburg[$], PhD (3)

(1) INSERM & Paris Diderot University, IAME, UMR 1137, Paris, France; (2), AP-HP, Bichat Hospital, Paris, France; (3) Da Volterra, Paris, France; (4) UNT Health Science Center, Fort Worth, TX, USA

[*, $]These authors contributed equally to the work

[§]Present affiliation: Vétoquinol, Paris, France

# Meeting presentations

# Funding

This work was supported by Da Volterra, Paris, France.

# Corresponding author

Jean de Gunzburg

Address: Da Volterra, Le Dorian Bât B1, 172 rue de Charonne, 75011 Paris, France

Mail: gunzburg@davolterra.com

Phone number: +44 7917 863 207

## Potential conflicts of interests

Charles Burdet and Thu Thuy Nguyen performed statistical work for the Da Volterra Company through a contract with INSERM UMR 1137.

Nathalie Saint-Lu and Sakina Sayah-Jeanne are employees of the Da Volterra Company. Christine Miossec is a previous employee of the Da Volterra Company.

Antoine Andremont is a scientific adviser of the Da Volterra Company within the framework of the French law on Innovation and Research.

France Mentré and Jean de Gunzburg are consultants for the Da Volterra Company.

# Abstract

Background

Lowering the gut exposure to antibiotics during treatments can prevent microbiota disruption. We evaluated the effect of an activated charcoal-based adsorbent, DAV131A, on fecal free moxifloxacin concentration and mortality in a hamster model of moxifloxacin-induced *C. difficile* infection.

Methods

215 hamsters receiving moxifloxacin subcutaneously ($D_1$-$D_5$) were orally infected at $D_3$ with *C. difficile* spores. They received various doses (0-1800mg/kg/day) and schedules (BID, TID) of DAV131A ($D_1$-$D_8$). Moxifloxacin concentration and *C. difficile* counts were determined at $D_3$, and mortality at $D_{12}$. We compared mortality, moxifloxacin concentration and *C. difficile* counts according to DAV131A regimens, and modelled the link between DAV131A regimen, moxifloxacin concentration and mortality.

Results

All hamsters that received no DAV131A died, but none of those that received 1800mg/kg/day. A significant dose-dependent relationship between DAV131A dose and (*i*) mortality rates, (*ii*) moxifloxacin concentration and (*iii*) *C. difficile* counts was evidenced. Mathematical modeling suggested that (*i*) lowering moxifloxacin concentration at $D_3$, which was 58µg/g (95%CI=50-66) without DAV131A, to 17µg/g (14-21) would reduce mortality by 90% and (*ii*) this would be achieved with a daily DAV131A dose of 703mg/kg (596-809).

Conclusions

In this model of *C. difficile* infection, DAV131A reduced mortality in a dose-dependent manner by decreasing fecal free moxifloxacin concentration.

# Main text

## Introduction

*Clostridium difficile* is a sporulating Gram positive bacillus that can lead to mild-to-severe intestinal infections, including pseudomembranous colitis and toxic megacolon (1). With 500,000 cases and 29,000 deaths in 2011 in the USA (2), the burden of *C. difficile* infection on the US healthcare system has reached $4.8 billion (3). *C. difficile* is the leading cause of healthcare-associated infections (4), and the Centers for Diseases Control consider it as an immediate public health threat (5).

Antibiotics are the main risk factors for *C. difficile* infections because the gut microbiota is exposed to high concentrations of the drugs during oral or parenteral treatments, resulting in its disruption (6, 7). Reducing this exposure thus appears appealing for limiting the consequences of antibiotic treatments on the microbiota. Such an approach has been pioneered by administering β-lactamases together with β-lactam antibiotics. This prevents colonization by resistant bacteria in mice (8, 9), as well as in dogs (10) and in humans (11). It also reduces antibiotic concentrations in the human gut (12, 13) during treatments, and preserves the intestinal microbiota in humanized gnotobiotic pigs (12). However, this promising approach is limited to β-lactams, when many other antibiotics are also at risk of provoking *C. difficile* infection, particularly fluoroquinolones (14).

In rats, delivering activated charcoal to the intestine allowed removing ciprofloxacin residues from the gut, and decreased antibiotic exposure of the microbiota without affecting its plasma pharmacokinetics (15). Similarly, we have shown that oral DAV131A, a charcoal-based adsorbent, decreased intestinal colonization by β-lactam resistant *Klebsiella pneumoniae* in cefotaxime-treated mice (16).

Here we used the Syrian hamster model of *C. difficile* infection, that recapitulates many aspects of the human infection (17) and has been widely used for evaluating new therapies against *C. difficile* infection (18-20), to assess the protective effect of DAV131A. We also developed a mathematical model to analyze the relationships between DAV131A regimens, fecal free moxifloxacin concentration and hamster mortality.

## Material and Methods

DAV131A

DAV131A is an activated charcoal-based adsorbent with high adsorption capacity (16). It was administered to hamsters by oral gavage after mixing with 0.25% w/v Natrosol® 250 Hydroxyethylcellulose. Hamsters from placebo groups received Natrosol® alone.

Hamster model of moxifloxacin-induced *C. difficile* infection

A previously developed hamster model of antibiotic-induced *C. difficile* infection was adapted to moxifloxacin (21). Male Golden Syrian hamsters (80-120 grams) received 30 mg/kg of moxifloxacin by the subcutaneous route at a time designated as $H_0$, once a day from day 1 ($D_1$) to day 5 ($D_5$). This dose was chosen as the lowest dose resulting in a 100% mortality rate in treated hamsters infected with *C. difficile* spores. It is not expected to cause any toxicity in hamsters since the minimal lethal intravenous dose reported in mice and rats is 100 mg/kg (22).

Animals were infected orally on day 3 ($D_3$), 4 hours after moxifloxacin administration ($H_4$), with $10^4$ spores of the non-epidemic *C. difficile* strain UNT103-1 (VA-11, REA J strain), TcdA+, TcdB+, cdtB–, vancomycin MIC = 2 µg/mL, moxifloxacin MIC = 16 µg/mL, clindamycin MIC > 256 µg/mL, ceftriaxone MIC = 128 µg/mL, obtained from Curtis Donskey, Ohio VA Medical Centre. All surviving hamsters were euthanized at day 12 ($D_{12}$). Animals were housed in conformity with NIH guidelines (23). All procedures were conducted at the University of North Texas Health Science Center in Fort Worth (Texas, USA) in accordance with Protocol 2012/13-21-A06 approved by the local Institutional Animal Care and Use Committee.

Studies

Three studies were conducted in order to test the protection afforded by DAV131A from lethal moxifloxacin-induced *C. difficile* infection. Their design are summarized in Table 1. All hamsters received moxifloxacin and were inoculated with *C. difficile* spores as described above. DAV131A was administered from $D_1$ to $D_8$.

In study 1, we aimed at analyzing the dose-response relationship between DAV131A daily dose and survival. To that end 4 groups of 10 hamsters each (groups 1C, 1E, 1G, 1I) were treated with increasing daily doses of DAV131 (200, 600, 1200 or 1800 mg/kg/day) administered *bis in die* (BID) 4 hours before ($H_{-4}$) and 1 hour after ($H_1$) moxifloxacin injection. Four groups of 10 hamsters each (groups 1B, 1D, 1F, 1H) received the same treatment plus an additional dose of DAV131A 10 hours before the first administration of moxifloxacin, *i.e.* at $D_1H_{-10}$. A control group receiving moxifloxacin alone (group 1A) was included.

In study 2, we compared the effect on survival of BID and *ter in die* (TID) administrations of a high dose of DAV131A, 1800 mg/kg/day. Two groups of 15 hamsters each received DAV131A at a dose of 600 mg/kg TID (at $H_{-4}$, *i.e.* 4 hours before, and at $H_1$ and $H_6$, *i.e.* respectively 1 and 6 hours after moxifloxacin administration, group 2B), or at a dose of 900 mg/kg BID (at $H_{-4}$ and $H_1$ as described above, group 2C). In addition, all these animals also received an additional initial dose of DAV131A at $D_1H_{-10}$. A control, untreated group (group 2A) was also included.

In study 3, we assessed the influence on survival of giving the first dose of DAV131A before ($H_{-4}$), concomitantly or after ($H_1$) the first antibiotic administration. Seven groups of 10 hamsters each were included, all receiving DAV131A BID at $H_{-4}$ and $H_1$ on $D_2$-$D_8$, but at specific timings on $D_1$. Two of these groups received 600 (group 3B) and 1200 (group 3E) mg/kg/day DAV131A, respectively, at $H_{-4}$ and $H_1$ on $D_1$; two other groups also received 600 (group 3C) and 1200 (group 3F) mg/kg/day of DAV131A but at $H_0$ and $H_5$ on $D_1$. Three groups received 600 (group 3D), 1200 (group 3G) and 1800 (group 3H) mg/kg/day of DAV131A, respectively, at $H_2$ and $H_7$ on $D_1$. The last group (group 3A) received DAV131A placebo with the same schedule as these last 3 groups.

Feces collection and analysis

We focused the analysis on $D_3$ which bracketed *C. difficile* inoculation. Two pools of feces were collected daily from $D_2$ to $D_4$ in all studies. The first was made of all pellets emitted in the first 12 hours after moxifloxacin administration ($H_0$-$H_{12}$ period) and the second was made of all pellets emitted in the period between 12 and 24 hours after moxifloxacin administration ($H_{12}$-$H_{24}$). As it is a natural and physiological behavior in hamsters, coprophagy was not controlled.

Fecal free moxifloxacin concentration was determined at $D_3$ on feces collected during the $H_0$-$H_{12}$ period. Fecal pools were stored at -80°C until performing the assay. On the day of the assay, feces were weighted, homogenized in sterile saline, and debris were eliminated by centrifugation. Fecal free moxifloxacin concentration was measured by microbiological assay (*B. subtilis* ATCC 6633) after incubation at 37°C for 24 hours (24), with a limit of quantification (LOQ) of 0.2 µg/g. Missing data were imputed according to the following algorithm: (*i*) if fecal free moxifloxacin concentration was available for the period $H_0$-$H_{12}$ at $D_2$ and $D_4$, the missing value was imputed to the arithmetic mean of these 2 values; (*ii*) if the concentration was known only for the period $H_0$-$H_{12}$ at $D_2$ or $D_4$, the missing value was imputed to this available value; (*iii*) otherwise, the missing data were not imputed and the animal was excluded from analysis. Data below the LOQ was imputed to the LOQ.

Fecal counts of *C. difficile* were determined extemporaneously at $D_3$ on the $H_{12}$-$H_{24}$ pool by plating serial dilutions of the samples on CDSA selective media (BBL *C. difficile* Selective agar, BD). Counts were read after anaerobic incubation at 37°C for 48h. Fecal counts below the LOQ (3.3 log10 CFU/g of feces) were imputed to the LOQ. Missing values were imputed using the counts at $D_4$ [$H_{12}$-$H_{24}$], if available. Otherwise, the missing counts were not imputed.

Statistical analysis

We compared mortality rates at $D_{12}$ in all hamsters according to DAV131A daily doses using the non-parametric Fisher exact test. The link between the DAV131A daily dose and (*i*) fecal free moxifloxacin concentration at $D_3$ [$H_0$-$H_{12}$] and (*ii*) the decimal logarithm of the *C. difficile* counts in feces at $D_3$ [$H_{12}$-$H_{24}$] was studied using the Spearman rank correlation test. Exact 95% confidence interval of the mortality rates were computed using the binomial distribution.

We compared fecal free moxifloxacin concentration and *C. difficile* counts according to vital status at $D_{12}$ using non-parametric Wilcoxon test. Ability of fecal free moxifloxacin concentration and *C. difficile* counts for predicting death were assessed using the area under the ROC curve and their 95% confidence interval, computed using 1000 paired-bootstrap replicates (R functions *roc* and *ci.auc*).

In hamsters who received DAV131A treatment (all groups except 1A, 2A and 3A), we compared fecal free moxifloxacin concentration according to the administration of an additional initial dose of DAV131A at $D_1H_{-10}$, or not using non-parametric Wilcoxon test.

The impact of the BID *vs* TID DAV131A administrations on fecal free moxifloxacin concentration was tested in hamsters receiving a daily dose of DAV131A of 1800 mg/kg that had received a dose of DAV131A at $D_1H_{-10}$ (groups 1H, 2B and 2C). The impact on fecal free moxifloxacin concentration of the timing of the first DAV131A administration (4 hours before, together with or 2 hours after the first moxifloxacin administration) was tested in hamsters receiving 600 or 1200 mg/kg/day of DAV131A who did not receive DAV131A at $D_1H_{-10}$ (groups 1E, 1G, 3B, 3C, 3D, 3E, 3F, 3G). Analyses were performed using non-parametric Wilcoxon or Kruskal-Wallis tests, as appropriate.

Finally, in order to identify independent features of the DAV131A dosing schedule associated with the reduction of fecal free moxifloxacin concentration, and to link the DAV131A dosing regimen to the mortality rate, we performed a modeling analysis of the data. Full methods and results are presented in Supplementary Text S1.

Data are presented as number of observations n (%) or median (min-max). All tests were 2-sided with a type-I error of 0.05. All analyses were performed using R software v3.2.2.

## Results

<u>Comparison of mortality, fecal free moxifloxacin concentration and *C. difficile* counts across DAV131A doses</u>

Values for fecal free moxifloxacin concentration at $D_3$ were missing for only 3/215 hamsters (1.4%) and 1 value was below the LOQ. Values for *C. difficile* counts at $D_3$ were missing for only 9/215 hamsters (4.2%), but as much as 121/215 (56.3%) were below the LOQ.

Descriptive statistics on fecal free moxifloxacin concentration, *C. difficile* counts and mortality rates in each group of each study are reported in Table 1.

All (100%, 95%CI=90.0-100) 35 hamsters from the control groups that received moxifloxacin but no DAV131A (groups 1A, 2A, 3A) died; they had a median fecal free moxifloxacin concentration of 53.8 µg/g (min-max, 24.5-211.3) at $D_3$ at the time of *C. difficile* inoculation. Conversely, none of the 60 animals receiving a daily dose of DAV131A of 1800 mg/kg (groups 1H, 1I, 2B, 2C, 3H) died (0%, 95%CI=0.0-6.0); they had a median fecal free moxifloxacin concentration of 1.8 µg/g only (min-max, 0.0 – 20.3). These animals showed no sign of disease, nor inflexion in their weight gain (data not shown). Also, the median counts of *C. difficile* were 6.0 log10 CFU/g (min-max, <3.3-7.8) in the control groups that received moxifloxacin but no DAV131A, much higher than the 3.8 log10 CFU/g (min-max, <3.3-5.9) observed in the moxifloxacin and 1800 mg/kg/day DAV131A treated group.

We observed a highly significant decrease in mortality when the DAV131A daily dose administered to hamsters increased ($p<10^{-15}$, Figure 1). A significant association was also evidenced between DAV131A daily dose and fecal free moxifloxacin concentration (Spearman rank correlation coefficient s=-0.9, $p<10^{-15}$, Figure 2) as well as between DAV131A daily dose and fecal counts of *C. difficile* (Spearman rank correlation coefficient s=-0.3, $p<10^{-4}$, Figure 3). Approximately 60% of the moxifloxacin excreted during $D_3$ was retrieved on the fecal pellets collected during the period $H_0$-$H_{12}$.

<u>Comparison of fecal free moxifloxacin concentration and *C. difficile* counts according to vital status.</u>

Median fecal concentration of free moxifloxacin was 46.0 µg/g (min-max, 12.3-463.4) in hamster in which death occurred by $D_{12}$ and 6.8 µg/g (min-max, 0.28-42.9) in hamsters that survived ($p<10^{-15}$). Similarly, *C. difficile* counts were higher in hamsters in which death occurred than in survivors (5.2 log10 CFU/g, min-max, <3.3-7.8 vs <3.3 log10 CFU/g, min-max, <3.3-5.9, $p<10^{-15}$). The areas under the ROC curves of the fecal concentration of free moxifloxacin and of *C. difficile* counts for predicting death were 0.97 (95%CI=0.95.-0.99) and 0.86 (95%CI=0.79-0.92), respectively.

<u>Influence of DAV131A dosing schedule on fecal free moxifloxacin concentration</u>

Among the 180 hamsters treated by DAV131A (all groups except 1A, 2A and 3A), the administration of an additional initial dose of DAV131A at $D_1H_{-10}$ was significantly associated with a lower fecal free moxifloxacin concentration (median [min-max]: 3.1 µg/g [0.0-463.4] *vs* 11.7 µg/g [0.7-62.9], $p<10^{-5}$, Supplementary Figure S1).

In the 40 hamsters treated with a daily dose of 1800 mg/kg DAV131A and receiving an additional initial dose of DAV131A at $D_1H_{-10}$ (groups 1H, 2B and 2C), fecal free moxifloxacin concentration was not significantly different between hamsters treated BID *vs* TID with DAV131A (1.6 µg/g [0.0-4.9] *vs* 1.9 µg/g [0.9-3.8], p=0.2, Supplementary Figure S2).

Among the 80 hamsters treated with a daily dose of 600 or 1200 mg/kg DAV131A BID and who did not receive an additional initial dose of DAV131A at $D_1H_{-10}$ (groups 1E, 1G, 3B, 3C, 3D, 3E, 3F, 3G), there was no significant difference in fecal free moxifloxacin concentration whether the first dose of DAV131A was administered 4 hours before (12.1 µg/g [5.1-27.8]), together with (15.6 µg/g [3.6-42.9]) or 2 hours after (12.8 µg/g [3.0-38.9]) the first administration of moxifloxacin (p=0.7, Supplementary Figure S3).

## Discussion

Our most important result was that DAV131A provided a dose-dependent reduction of mortality in a hamster model of moxifloxacin-induced *C. difficile* infection. Fecal free moxifloxacin concentration of 53.8 µg/g (min-max, 24.5-211.3), *C. difficile* counts of 6.0 log10 CFU/g (min-max, 3.1-7.8) and a 100% mortality rate were observed in animals who did not receive DAV131A treatment, while fecal free moxifloxacin concentration, *C. difficile* counts and mortality were respectively decreased to 7.3 µg/g (min-max, 3.0-29.6), 3.8 log10 CFU/g (min-max, 3.1-5.9) and 0% with doses of 1800 mg/kg/day. DAV131A is the first product to exhibit such a level of protection against mortality from antibiotic-induced *C. difficile* infection in hamsters. Indeed a polymeric toxin-binding compound had been shown to protect only 70% to 90% of hamsters in an animal model of clindamycin-induced *C. difficile* infection (25), however, 20% to 40% of animals from the toxin-binding treatment group still had diarrhea 15 days after cessation of therapy, whereas protected animals in the experiments reported here had no sign of disease (data not shown).

The schedule of DAV131A administration (4 hours before, together or 2 hours after the first moxifloxacin administration) did not significantly affect the protective effect as assessed by survival, fecal free moxifloxacin concentration and *C. difficile* counts in feces (Table 1 and Supplementary Figure S3). When comparing administration schedules for a same total daily dose of DAV131A, BID was found not to be significantly less protective than TID. However, this result was only drawn from analysis of only 15 hamsters that received DAV131A on a TID-basis, all of which were treated with DAV131A at the highest dose (1800 mg/kg/day) and had also received an additional initial dose of DAV131A at $D_1H_{-10}$.

Another important result of our work is that the modelling approach described in the Supplementary Text S1 section allowed to investigate the mechanism of action by which DAV131A reduced mortality in hamsters. The effect appeared mediated by the reduction of the fecal concentration of free moxifloxacin when increasing the dose of DAV131A, in a dose-dependent manner.

Our results should however be tempered by the absence of bacteriology data from our modeling analysis. We did not include *C. difficile* counts in the model as their ability to predict death was lower than that of fecal concentrations of free moxifloxacin. Furthermore, the symptoms of *C. difficile* infection are related to the action of toxins produced by pathogenic strains of *C. difficile* (1), whose presence and activity could not be assessed from fecal samples in our studies. We are currently developing new methods for measuring toxin production and activity in order to perform a more thorough analysis of the protective effect of DAV131A.

Altogether, our data provide encouraging prospects for the protection of the gut microbiota from perturbation during the use of antimicrobials, such as fluoroquinolones which are widely used for therapeutic purposes and were associated with the rise of the hypervirulent epidemic *C. difficile* strains of the 027 ribotype (14). These results in hamsters suggest that the approach warrants further clinical development. Indeed, the hamster model of *C. difficile* infection is appropriate for reproducing the deleterious impact of antibiotic treatments on gut microbiota that allows *C. difficile* spores to germinate and the disease to develop (21). Therefore we believe that the efficacy of DAV131A obtained in this model is also relevant with respect to the mechanism of action of the product. The modelling approach confirms this, by showing that the protective effect of DAV131A in the hamster model was mediated by the reduction of the antibiotic concentrations in the gut.

The interpolation of the dose of adsorbent between hamsters and humans constitutes a challenge, because of the vast differences in gastrointestinal transit times and fecal excretion physiology between the two species. Additionally, they could also exhibit differences in antibiotic pharmacokinetics; however, the fact that the maximum fecal concentration of free moxifloxacin measured in hamsters in the experiments reported here were within the range of what was found in humans treated with a clinical dose of moxifloxacin suggests similarities in the selective pressure exerted by antibiotics on the intestinal microbiota of hamsters in the model developed in this study and humans (26). Even given the limitations discussed above in transposing results between animal models and human patients, in particular because gastrointestinal transit is much faster in hamsters than in humans, our results suggest that the product would not necessarily need to be administered before the antibiotic, but could be given concomitantly or just after the first antibiotic intake. In the case where antimicrobial therapy can be programmed, a supplementary protection might be obtained when pre-treating patients with DAV132, as suggested by the additional protection obtained in hamsters which received the first dose of DAV131A 10 hours before the first injection of moxifloxacin. This should be further investigated in human studies.

It can thus be inferred from our results that DAV131A protected the animals against *C. difficile* colonization and infection by preventing the disruption of the gut bacterial microbiota which is known to occur during fluoroquinolone treatments (6) and which constitutes the primary risk factor for *C. difficile* infection in humans (7). Therefore DAV132, the human counterpart of DAV131A that has recently been developed and tested in human volunteers (27), could represent a promising approach for the prevention of *C. difficile* infection during antibiotic treatments. A first phase 1 clinical trial in human volunteers has recently shown that the active component of DAV131A contained in the human-directed product DAV132 could be targeted to the ileo-caecal region in humans, and that its concomitant use with an orally-administered antibiotic did not impact the plasma pharmacokinetics of the antibiotic (27). Studies to expand these results to patients and establish the efficacy of DAV132 to prevent *C. difficile infections* following antibiotic treatments are currently underway.

# Figures

Figure 1. Mortality rates at $D_{12}$ according to DAV131A daily doses administered to the 215 hamsters of the 3 pooled studies. Bars represent the exact 95% confidence intervals of observed proportions.

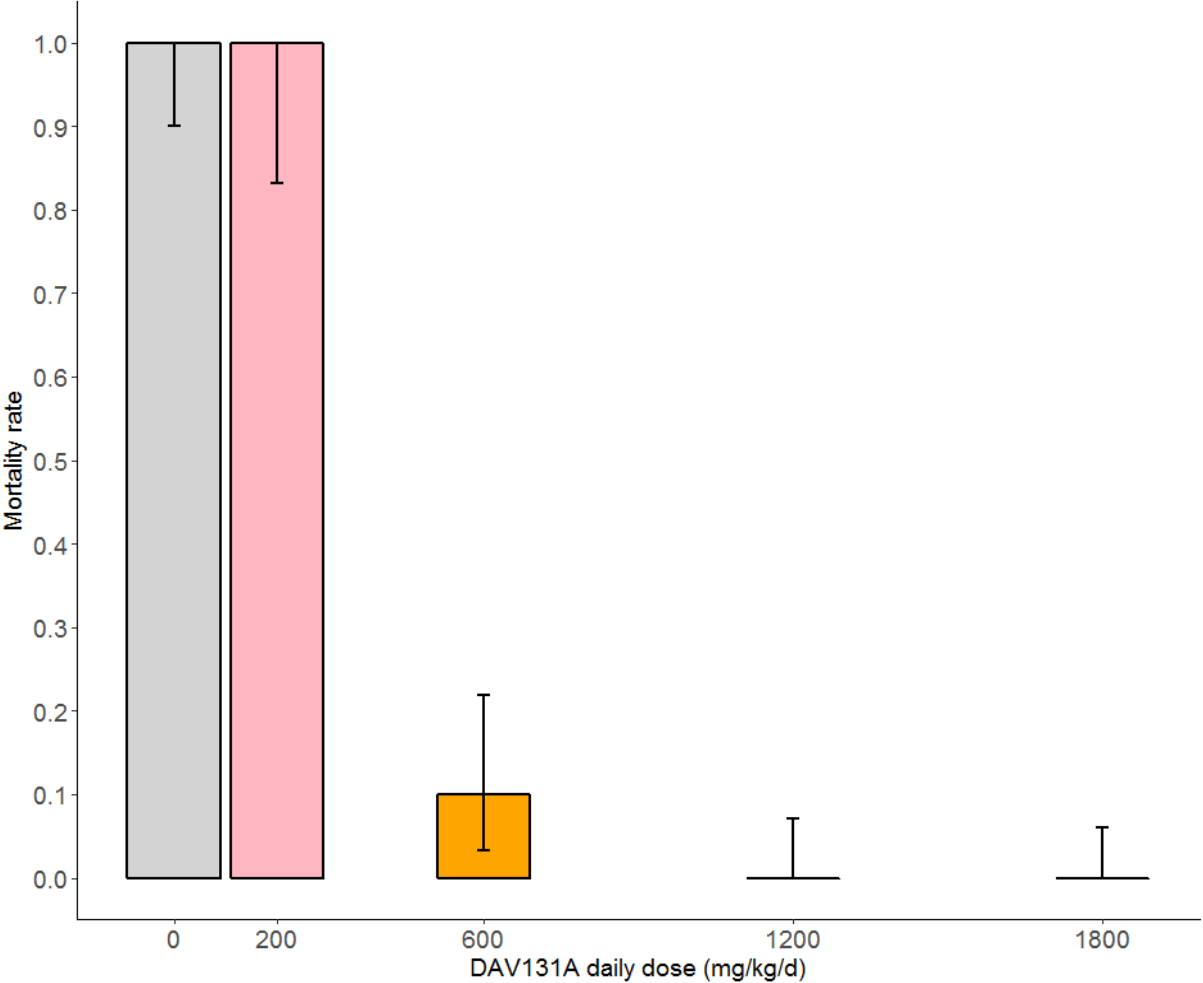

Figure 2. Boxplots of the fecal free moxifloxacin concentration measured at $D_3$ according to the DAV131A daily dose administered in the 212 hamsters of the 3 pooled studies. Triangles, dots and squares represent the observed concentrations in studies 1, 2 and 3, respectively. Whiskers represent $10^{th}$ and $90^{th}$ percentiles. Red symbol represents data below the limit of quantification.

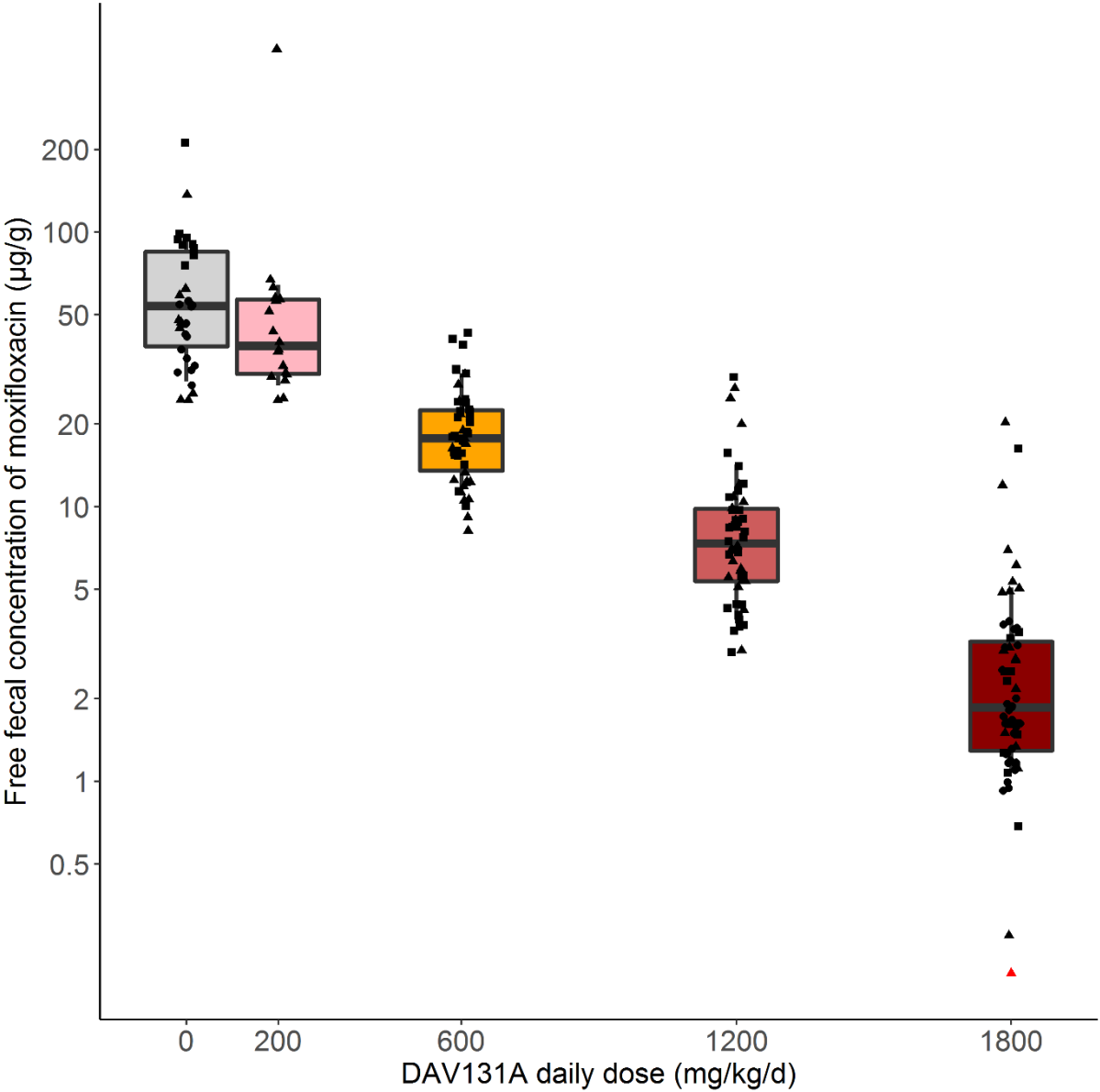

Figure 3. Boxplots of the counts of *Clostridium difficile* measured at $D_3$ according to the DAV131A daily dose administered in the 206 hamsters of the 3 pooled studies. Triangles, dots and squares represent the observed concentrations in studies 1, 2 and 3, respectively. Whiskers represent 10th and 90th percentiles. Red symbols represent data below the limit of quantification.

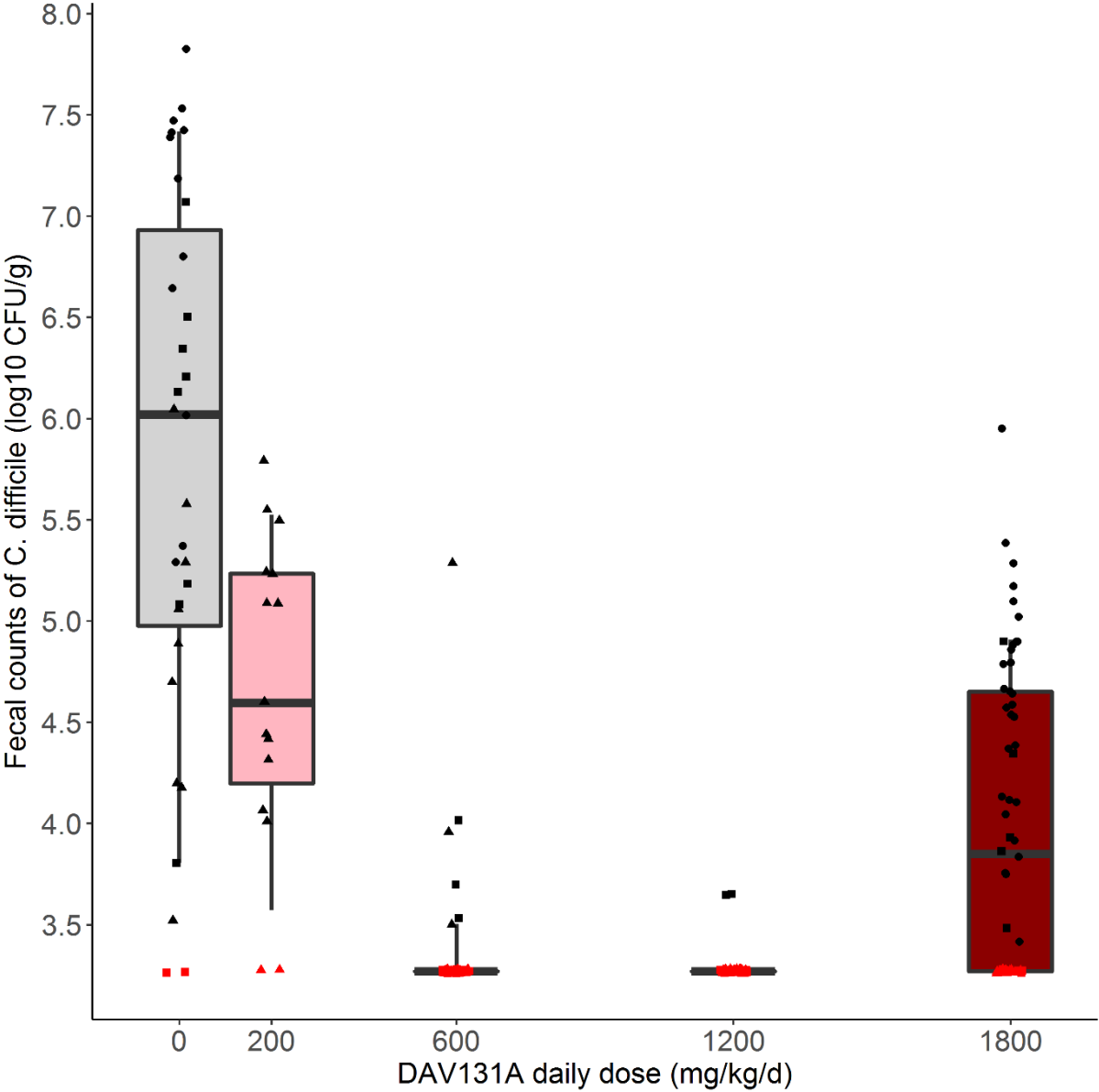

# Tables

Table 1. Descriptive statistics on fecal free moxifloxacin concentration at $D_3$, *Clostridium difficile* log10-counts at $D_3$ and mortality rates at $D_{12}$ according to DAV131A daily dose in the three studies.

| Study | Group number | DAV131A administration | | | | | Fecal free moxifloxacin concentration at $D_3$ [$H_0$-$H_{12}$] (µg/g) | | | | | *C. diff* log10-counts at $D_3$ [$H_{12}$-$H_{24}$] (log10 CFU/g) | | | | Mortality at $D_{12}$ |
|---|---|---|---|---|---|---|---|---|---|---|---|---|---|---|---|---|
| | | Daily dose (mg/kg/day) | Administration at $D_1H_{-10}$ | Dosing time at $D_1$ | Dosing time at $D_2$-$D_8$ | N | N | Median | Min | Max | N | Median | Min | Max | N (%) |
| study 1 ($N_1$=90) | 1A | 0 (no placebo) | no | - | - | 10 | 10 | 46.4 | 24.5 | 136.5 | 9 | 4.9 | 3.5 | 6.0 | 10 (100) |
| | 1B | 200 (100 BID) | yes | $H_{-4}, H_1$ | $H_{-4}, H_1$ | 10 | 10 | 47.5 | 24.5 | 463.4 | 8 | 4.8 | <3.3 | 5.8 | 10 (100) |
| | 1C | 200 (100 BID) | no | $H_{-4}, H_1$ | $H_{-4}, H_1$ | 10 | 8 | 34.6 | 24.8 | 62.9 | 7 | 4.4 | 4 | 5.5 | 10 (100) |
| | 1D | 600 (300 BID) | yes | $H_{-4}, H_1$ | $H_{-4}, H_1$ | 10 | 10 | 14.8 | 8.2 | 30.4 | 10 | <3.3 | <3.3 | <3.3 | 0 (0) |
| | 1E | 600 (300 BID) | no | $H_{-4}, H_1$ | $H_{-4}, H_1$ | 10 | 10 | 12.3 | 9.1 | 27.8 | 10 | <3.3 | <3.3 | <3.3 | 2 (20) |
| | 1F | 1200 (600 BID) | yes | $H_{-4}, H_1$ | $H_{-4}, H_1$ | 10 | 10 | 7.1 | 3.0 | 10.8 | 10 | <3.3 | <3.3 | <3.3 | 0 (0) |
| | 1G | 1200 (600 BID) | no | $H_{-4}, H_1$ | $H_{-4}, H_1$ | 10 | 10 | 6.6 | 5.1 | 27 | 10 | <3.3 | <3.3 | <3.3 | 0 (0) |
| | 1H | 1800 (900 BID) | yes | $H_{-4}, H_1$ | $H_{-4}, H_1$ | 10 | 10 | 1.5 | 0.2 | 4.9 | 10 | <3.3 | <3.3 | <3.3 | 0 (0) |
| | 1I | 1800 (900 BID) | no | $H_{-4}, H_1$ | $H_{-4}, H_1$ | 10 | 10 | 5.2 | 1.6 | 20.3 | 10 | <3.3 | <3.3 | <3.3 | 0 (0) |
| study 2 ($N_2$=45) | 2A | 0 (no placebo) | no | - | - | 15 | 14 | 42.0 | 27.7 | 56.2 | 12 | 7.3 | 5.3 | 7.8 | 15 (100) |
| | 2B | 1800 (600 TID) | yes | $H_{-4}, H_1, H_6$ | $H_{-4}, H_1, H_6$ | 15 | 15 | 1.6 | 1.0 | 2.0 | 15 | 4.6 | 3.4 | 5.4 | 0 (0) |
| | 2C | 1800 (900 BID) | yes | $H_{-4}, H_1$ | $H_{-4}, H_1$ | 15 | 15 | 1.9 | 0.9 | 3.8 | 15 | 4.6 | 3.7 | 5.9 | 0 (0) |
| study 3 ($N_3$=80) | 3A | 0 (placebo) | no | - | - | 10 | 10 | 90 | 75.5 | 211.3 | 10 | 5.7 | <3.3 | 7.1 | 10 (100) |
| | 3B | 600 (300 BID) | no | $H_{-4}, H_1$ | $H_{-4}, H_1$ | 10 | 10 | 19.6 | 10 | 24.1 | 10 | <3.3 | <3.3 | 4.0 | 1 (10) |
| | 3C | 600 (300 BID) | no | $H_0, H_5$ | $H_{-4}, H_1$ | 10 | 10 | 23.4 | 17.3 | 42.9 | 10 | <3.3 | <3.3 | 3.7 | 0 (0) |
| | 3D | 600 (300 BID) | no | $H_2, H_7$ | $H_{-4}, H_1$ | 10 | 10 | 15.7 | 11.3 | 38.9 | 10 | <3.3 | <3.3 | <3.3 | 2 (20) |
| | 3E | 1200 (600 BID) | no | $H_{-4}, H_1$ | $H_{-4}, H_1$ | 10 | 10 | 8.9 | 5.6 | 15.7 | 10 | <3.3 | <3.3 | <3.3 | 0 (0) |
| | 3F | 1200 (600 BID) | no | $H_0, H_5$ | $H_{-4}, H_1$ | 10 | 10 | 5 | 3.6 | 14 | 10 | <3.3 | <3.3 | <3.3 | 0 (0) |
| | 3G | 1200 (600 BID) | no | $H_2, H_7$ | $H_{-4}, H_1$ | 10 | 10 | 7.3 | 3.0 | 29.6 | 10 | <3.3 | <3.3 | 3.7 | 0 (0) |
| | 3H | 1800 (900 BID) | no | $H_2, H_7$ | $H_{-4}, H_1$ | 10 | 10 | 2.4 | 0.7 | 16.3 | 10 | <3.3 | <3.3 | 4.9 | 0 (0) |
| All groups | | | | | | 215 | 212 | 11.6 | 0.2 | 463.4 | 206 | <3.3 | <3.3 | 7.8 | 60 (27.9) |